

\magnification=\magstep1       	
\font\bigbold=cmbx10 scaled 1200
\let\frak=\cal			

\newcount\EQNO      \EQNO=0
\newcount\FIGNO     \FIGNO=0
\newcount\REFNO     \REFNO=0
\newcount\SECNO     \SECNO=0
\newcount\SUBSECNO  \SUBSECNO=0
\newcount\FOOTNO    \FOOTNO=0
\newbox\FIGBOX      \setbox\FIGBOX=\vbox{}
\newbox\REFBOX      \setbox\REFBOX=\vbox{}
\newbox\RefBoxOne   \setbox\RefBoxOne=\vbox{}

\expandafter\ifx\csname normal\endcsname\relax\def\normal{\null}\fi

\def\Eqno{\global\advance\EQNO by 1 \eqno(\the\EQNO)%
    \gdef\label##1{\xdef##1{\nobreak(\the\EQNO)}}}
\def\Fig#1{\global\advance\FIGNO by 1 Figure~\the\FIGNO%
    \global\setbox\FIGBOX=\vbox{\unvcopy\FIGBOX
      \narrower\smallskip\item{\bf Figure \the\FIGNO~~}#1}}
\def\Ref#1{\global\advance\REFNO by 1 \nobreak[\the\REFNO]%
    \global\setbox\REFBOX=\vbox{\unvcopy\REFBOX\normal
      \smallskip\item{\the\REFNO .~}#1}%
    \gdef\label##1{\xdef##1{\nobreak[\the\REFNO]}}}
\def\Section#1{\SUBSECNO=0\advance\SECNO by 1
    \bigskip\leftline{\bf \the\SECNO .\ #1}\nobreak}
\def\Subsection#1{\advance\SUBSECNO by 1
    \medskip\leftline{\bf \ifcase\SUBSECNO\or
    a\or b\or c\or d\or e\or f\or g\or h\or i\or j\or k\or l\or m\or n\fi
    )\ #1}\nobreak}
\def\Footnote#1{\global\advance\FOOTNO by 1
    \footnote{\nobreak$\>\!{}^{\the\FOOTNO}\>\!$}{#1}}
\def\SameFootnote{$\>\!{}^{\the\FOOTNO}\>\!$}

\def\References{\bigskip\centerline{\bf REFERENCES}
                \smallskip\copy\REFBOX}
\def\NewRefPage{\setbox\RefBoxOne=\vbox{\unvcopy\REFBOX}%
		\setbox\REFBOX=\vbox{}%
		\def\References{\bigskip\centerline{\bf REFERENCES}
                		\nobreak\smallskip\nobreak\copy\RefBoxOne
				\vfill\eject
				\smallskip\copy\REFBOX}%
		\def\NewRefPage{}}






\def\Fig#1{\global\advance\FIGNO by 1 Figure~\the\FIGNO: {#1}%
    \global\setbox\FIGBOX=\vbox{\unvcopy\FIGBOX
      \narrower\smallskip\item{\bf Figure \the\FIGNO~~}#1}}

\newcount\ITEMNO     \ITEMNO=0

\def\itemno{\global\advance\ITEMNO by 1 \the\ITEMNO}


\long\def\Theorem#1#2{\medbreak\noindent{\bf Theorem \itemno
\enspace}{\sl#1}\par\medbreak {\medbreak\narrower {\it Proof:}
{#2}\hfill{$\spadesuit$}\medbreak\smallskip}}

\long\def\Lemma#1#2{\medbreak\noindent{\bf Lemma \itemno
\enspace}{\sl#1}\par\medbreak {\medbreak\narrower {\it Proof:}
{#2}\hfill{$\spadesuit$}\medbreak\smallskip}}

\def\Theoremwoproof#1{\medbreak\noindent{\bf Theorem \itemno
\enspace}{\sl#1}\par\medbreak}

\def\Definition#1{\medbreak\noindent{\bf Definition \itemno
\enspace}{\sl#1}\par\medbreak}

\def\Proposition#1#2{\medbreak\noindent{\bf Proposition \itemno
\enspace}{\sl#1}\par\medbreak {\medbreak\narrower {\it Proof:} #2\
$\spadesuit$\medbreak\smallskip}}

\vsize=9truein
\nopagenumbers			
\headline={\ifnum\pageno=1{\hss}\else{\hss\rm -~\folio~- \hss}\fi}


\def\cross{\times}


\def\Gamu#1#2#3{\Gamma^{#1}_{\ #2#3}}

\def\Gamperpu#1#2#3{{^\perp}\!\Gamma^{#1}_{\ #2#3}}


\def\frac#1#2{{{#1}\over {#2}}}

\def\at#1{\lower.8ex\hbox{${\big\vert _{#1}}$}}

\def\comma#1{{_{,#1}}}
\def\starry#1{{_{*#1}}}
\def\pel#1{{\rm P}_i}
\def\dt{\frac{\partial}{\partial t}}

\def\dxi{\frac{\partial}{\partial x^i}}

\def\dxalpha{\frac{\partial}{\partial x^\alpha}}

\def\partiali{\partial_i}

\def\partiall{\partial_l}

\def\partialt{\partial_t}

\def\partialalpha{\partial_\alpha}
\def\partialbeta{\partial_\beta}
\def\partialgamma{\partial_\gamma}
\def\spartiali{\partial_{*i}}
\def\spartialj{\partial_{*j}}
\def\spartialk{\partial_{*k}}
\def\spartiall{\partial_{*l}}

\def\l#1{\hbox{\it \$}_{\!\scriptscriptstyle#1}}	
\def\lstar#1{\hbox{\it
\$}_{\!\!*\,\,\lower2pt\hbox{$\!\!\scriptscriptstyle#1$}}}

\def\del#1{{\nabla_{#1}}}

\def\sdel#1{{\nabla_{\!\!*#1}}}

\def\d#1{D_{#1}}

\def\riemu#1#2#3#4{R^{#1}_{\ #2#3#4}}

\def\rperpu#1#2#3#4{{^\perp\! R}^{#1}_{\ #2#3#4}}
\def\rbarperpu#1#2#3#4{{^\perp\!\bar R}^{#1}_{\ #2#3#4}}
\def\rfour{{^4\!R}}
\def\rthree{{^3\!R}}
\def\rperp{{^\perp\!R}}
\def\rbarperp{{^\perp\!\bar R}}
\def\rhat{\hat R}
\def\zelmanovu#1#2#3#4{Z^{#1}_{\ #2#3#4}}

\def\D{{\cal D}}				
\def\tor#1#2#3{{\rm T}^{#1}_{\ #2#3}}
\def\tord{T_{\!D}}
\def\tordperp{^\perp\!T_{\!D}}
\def\tornabla{T_{\!\nabla}}

\def\kdelta#1#2{\delta^{#1}_{\ #2}}
\def\downkdelta#1#2{\delta_{#1}^{\ #2}}
\def\proj#1#2{P_{#1}^{\ #2}}

\def\sqr#1#2{{\vbox{\hrule height.#2pt
              \hbox{\vrule width.#2pt height#1pt \kern#1pt
                   \vrule width.#2pt}\hrule height.#2pt}}}


\def\hij{h_{ij}}
\def\gij{g_{ij}}

\def\metric#1#2{\left< #1 , #2 \right>}
\def\bracket#1#2{\left[ #1 , #2 \right]}

\def\p1tensors{{\cal P}\! T_1}



\def\tpsigmaperp{(T_p\Sigma)^\perp}
\def\tpmperp{(T_p\M)^\perp}
\def\tmperp{(T\M)^\perp}
\def\llbrak{\left<\!\left<}
\def\rrbrak{\right>\!\right>}
\def\vfs{\raise3pt\hbox{$\chi$}(\M)}
\def\pvfs{\raise3pt\hbox{$\chi_{_*}$}(\Sigma)}
\def\fcns{{\frak F}(\M)}		



\def\j1pi{J^1_{\pi}}

\def\hforms{\bigwedge\,\raise6pt\hbox{$\!\!\scriptstyle
1$}\lower6pt\hbox{$\!\!\scriptstyle 0$} \pi}

\def\M{{\cal M}} 		


\def\tornabla{T}
\def\rfour{{R}}
\def\rthree{{R_{\!D}}}
\def\dt{\partialt}
\def\dxi{\partiali}
\def\dxalpha{\partialalpha}
\def\perpvfs{\raise3pt\hbox{$\chi$}^\perp}


\def\today{\number\day\space\ifcase\month\or
  January\or February\or March\or April\or May\or June\or
  July\or August\or September\or October\or November\or December\fi
  \space\number\year}
\rightline{12 July 1994}
\rightline{gr-qc/9407011}
\bigskip\bigskip

\null\bigskip\bigskip\bigskip
\centerline{\bigbold PARAMETRIC MANIFOLDS I: Extrinsic Approach}
\bigskip

\centerline{Stuart Boersma}
\centerline{\it Department of Mathematics, Oregon State University,
		Corvallis, OR  97331, USA
\Footnote{Present address: Division of Mathematics and Computer Science,
Alfred University, Alfred, NY  14802}
}
\centerline{\tt boersma@math.orst.edu}
\medskip
\centerline{Tevian Dray}
\centerline{\it Department of Mathematics, Oregon State University,
		Corvallis, OR  97331, USA}
\centerline{\tt tevian@math.orst.edu}

\bigskip\bigskip\bigskip\bigskip
\centerline{\bf ABSTRACT}
\midinsert

\narrower\narrower\noindent
A {\it parametric manifold} can be viewed as the manifold of orbits of a
(regular) foliation of a manifold by means of a family of curves.  If the
foliation is hypersurface orthogonal, the parametric manifold is equivalent to
the 1-parameter family of hypersurfaces orthogonal to the curves, each of
which inherits a metric and connection from the original manifold via
orthogonal projections; this is the well-known Gauss-Codazzi formalism.  We
generalize this formalism to the case where the foliation is not hypersurface
orthogonal.  Crucial to this generalization is the notion of {\it deficiency},
which measures the failure of the orthogonal tangent spaces to be
surface-forming, and which behaves very much like torsion.  Some applications
to initial value problems in general relativity will be briefly discussed.

\endinsert
\vfill
\eject


\Section{Introduction}

Associated with a foliation of spacetime by spacelike hypersurfaces is the
dual foliation by timelike curves orthogonal to the hypersurfaces, {\it i.e.}\
the trajectories of observers whose instantaneous rest spaces consist
precisely of the given hypersurfaces.  But how does one describe physics as
seen by observers whose trajectories are {\bf not} hypersurface orthogonal?
It is the goal of this paper to describe one possible framework for answering
such questions.

The decomposition of various fields on a manifold into data on a hypersurface
is not merely of interest for spacetimes.  Given a (non-degenerate) metric of
any signature on a manifold $\M$, the Gauss-Codazzi equations show how to
project the geometry of $\M$ orthogonally onto a hypersurface $\Sigma$.  This
paper generalizes the Gauss-Codazzi equations which describe the geometry
orthogonal to a given family of curves to include the case when these curves
fail to be hypersurface-orthogonal.

The term {\it parametric manifold} has been recently coined by Perj\'es
\Ref{{Z. Perj\'es}, {\it The Parametric Manifold Picture of Space-Time},
{Nuclear Physics} {\bf B403}, 809 (1993)} \label\PERJES
in this setting.  He traces some of the geometric ideas back to Zel'manov
\Ref{{A. Zel'manov}, {Soviet Physics Doklady} {\bf 1}, 227--230
(1956)}\label\ZELMANOV
; similar ideas can also be found in some work of Einstein and Bergmann
\Ref{{A. Einstein and P. Bergmann}, {Ann.~of Math} {\bf 39}, 683--701, (1938).
\hfill\break
{\it This article was reprinted in}
{\bf Introduction to Modern Kaluza-Klein Theories}, edited by T.
Applequist, A. Chodos, and P.G.O. Freund, Addison-Welsey, Menlo Park, 1987.}
\label\EIN
on Kaluza-Klein theories.  However, none of these authors describe the
parametric theory in modern mathematical language, as tensors are given in
terms of their components in a coordinate basis and their abstract properties
are not clear.  In particular, defining {\it torsion} in this setting, and
especially distinguishing it from the new concept of {\it deficiency}, is
hard to do without a basis-free approach.  This paper presents one way of
unifying these earlier ideas into a rigorous mathematical framework.

We start by reviewing some basic properties of connections in Section 2,
followed by a description of the usual Gauss-Codazzi formalism in Section 3.
We have deliberately presented some fairly standard material in considerable
detail so that the comparison with the generalized Gauss-Codazzi framework,
described in Section 4, will be clear.  In Section 5, we express our results
in a coordinate basis so that it can be more easily compared to earlier work.
In Section 6 we then show that our framework does {\bf not}, in fact,
completely reproduce the earlier results cited above, and we further show how
this can be remedied.  Finally, in Section 7, we discuss our results.


\Section{Background}

Let us begin by reviewing the standard notion of a connection on a manifold
$\M$, together with some relevant properties of connections.  For the
following definitions, let $\M$ be a smooth manifold with (Lorentzian or
Riemannian) metric $g$ denoted by $\langle\ ,\ \rangle$.  Also, let $\vfs$
denote the set of all smooth vector fields on $\M$ and $\fcns$ the ring of all
smooth real-valued functions defined on $\M$.

\Definition{An (affine) connection $\nabla$ on $\M$ is a mapping $\nabla :
\vfs \cross \vfs \to \vfs$, usually denoted by $\nabla (X,Y) = \del X Y$,
which satisfies the following axioms:
\item{i.} Linearity over $\fcns$: $\del{fX+gY}Z = f\del X Z + g\del Y Z$
\item{ii.} Linearity: $\del X (Y+Z) = \del XY + \del XZ$
\item{iii.} Product rule: $\del X (fY) = f \del XY + X(f)\, Y$ for all
$X,Y,Z\in\vfs$ and $f,g\in\fcns$.}

The existence of a connection on $\M$ provides a way of differentiating vector
fields along curves, which can be extended in the usual way to be a derivation
on all tensor fields.  Although traditionally one defines the concept of metric
compatibility in terms of parallel vector fields along curves in $\M$, it can
be restated ({\it cf.}\
\Ref{{M. do Carmo}, {\bf Riemannian Geometry}, {translated by F. Flaherty,
Birkh\"auser, Boston, 1992.}}\label\DOCARMO
) as

\Definition{An affine connection $\nabla$ is compatible with the metric of
$\M$ provided
$$X\Big(\metric YZ\Big) = \metric{\del XY}Z + \metric Y{\del XZ}
  \Eqno$$\label\DMetComp
for $X,Y,Z\in\vfs$.}

\Definition{A connection $\nabla$ is said to be torsion-free when
$$\del XY - \del YX = \bracket XY$$
for all $X,Y\in\vfs$.}

The action of $\bracket XY$ on functions $f\in\fcns$ is defined by the action
of the commutator
$$\bracket XY f = XYf - YXf.\Eqno$$\label\DBrak
Although it is not {\it a priori} clear that with this definition $\bracket
XY$ is a vector field, it can be shown ({\it cf.}\
\Ref{{R. Bishop and S. Goldberg}, {\bf Tensor Analysis On Manifolds}, {Dover
Publications, Inc., New York, 1980.}}\label\BandG
) that there exists a unique vector field, also written $\bracket XY$,
satisfying \DBrak .

A fundamental result in the theory of connections is

\Theoremwoproof{There exists a unique connection on $\M$ which is compatible
with the metric $g$ and torsion-free.}

\Definition{This unique connection is called the Levi-Civita connection.}

The Levi-Civita connection can be given explicitly as ({\it e.g.}\
\Ref{{B. O'Neill}, {\bf Semi-Riemannian Geometry with Applications to
Relativity}, {Academic Press, Orlando, 1983}}\label\ONEILL
)
$$\eqalign{\metric Z{\del YX}&=\frac 12 \bigg( X\Big(\metric YZ\Big)
+ Y\Big(\metric ZX\Big) - Z\Big(\metric XY\Big)\cr
&+\metric{\bracket ZX}Y - \metric{\bracket YZ}X - \metric{\bracket
XY}Z\bigg).\cr}$$


The notions of curvature and torsion play an interesting role in the
development of a parametric theory.  A clear understanding of the
relationships between them will be useful when defining parametric curvature.
Using the definitions in
\Ref{{S. Kobayashi and K. Nomizu}, {\bf Foundations of Differential Geometry,
vol.\ I}, {John Wiley \& Sons, New York, 1963.}}\label\KOBAYASHI
, rewritten in terms of an affine
connection, we have

\Definition{The torsion $T$ and curvature $R$ of $\nabla$ are given by
$$T(X,Y) = \del XY - \del YX - \bracket XY \Eqno$$\label\DefT
and
$$R(X,Y)Z = \del X \del Y Z - \del Y \del X Z - \del{\bracket XY} Z
\Eqno$$\label\DefR
for $X,Y,Z\in\vfs$.}
The case where $T(X,Y)\equiv 0$ agrees with the earlier notion of
torsion-free.

Consider the components of $T$ and $R$ in some patch with coordinates
$\{x^\alpha\}$, so that the coordinate vector fields $\{\partialalpha\}$ form
a (local) basis of $\vfs$.  Defining the Christoffel symbols $\Gamu
\alpha\beta\gamma$ by
$\del{\partialbeta}\partialgamma = \Gamu \alpha\beta\gamma \partialalpha$
we have
$$\eqalign{T(\partialbeta,\partialgamma) &= \del\partialbeta \partialgamma -
\del\partialgamma \partialbeta - 0 \cr
	&=(\Gamu \alpha\beta\gamma - \Gamu\alpha\gamma\beta)\partialalpha\cr
	&=\tor\alpha\beta\gamma \partialalpha.\cr}$$

A torsion-free connection is thus sometimes referred to as a {\it symmetric
connection}.  While it is trivially true that mixed partial derivatives
commute, the torsion tensor may be thought of as measuring the failure of
mixed covariant derivatives to commute.  As we see from above
$$\left(\del\partialbeta \del\partialgamma - \del\partialgamma \del\partialbeta
\right) (f) = \tor \alpha\beta\gamma \partialalpha f$$
where we have used the fact that
$\del\partialalpha f=\partialalpha f={\partial f \over \partial x^\alpha}$

For curvature,
$$\eqalign{R(\partialalpha ,\partialbeta)\partialgamma &=(\del\partialalpha
\del\partialbeta - \del\partialbeta \del\partialalpha)\partialgamma - 0\cr
	&= \riemu\mu\gamma\alpha\beta \partial_\mu .\cr}\Eqno$$\label\Rcomp

As is often done, $\riemu \nu\delta\beta\alpha$ may be expressed in terms of
the Christoffel symbols $\Gamu \alpha\beta\gamma$,
$$\riemu \nu\delta\beta\alpha =
	  {\partial\Gamu\mu\delta\alpha \over \partial x^\beta}
	- {\partial\Gamu\mu\delta\beta \over \partial x^\alpha}
	+ \Gamu\nu\mu\beta \Gamu \mu\delta\alpha
	- \Gamu \nu\mu\alpha \Gamu\mu\delta\beta.$$

It is worth noting that in the definition \DefR\ of $R$, as well as in the
formula \Rcomp\ for the components $\riemu\mu\gamma\alpha\beta$, there is no
explicit mention of the torsion.  The case is different when using the
abstract index notation (see
\Ref{{R. Wald}, {\bf General Relativity}, {The University of Chicago Press,
Chicago, 1984}}\label\WALD
), which closely parallels component
notation in a coordinate basis.

In the abstract index notation, the vector field $\del XY$ is represented by
$X^a \del a Y^b$.  In a coordinate basis (with coordinates
$\{x^\alpha\}$), $X^a$ is the vector field $X^\alpha \partialalpha$.
Furthermore, in this notation
$\del a Z^b = \partial_a Z^b + \Gamu bca Z^c$
would represent the tensor with components
$${\partial Z^\beta\over\partial x^\alpha} + \Gamu\beta\gamma\alpha Z^\gamma
.$$

In the absence of torsion, one often defines the action of the Riemann
curvature tensor by
$$\riemu ncba Z_n = (\del a \del b - \del b \del a)Z_c.$$
However, in terms of the Christoffel symbols $\Gamu bca$
$$\eqalign{\del a \del b Z_c =\,&\partial_a\left(\partial_b Z_c
	- \Gamu mcb Z_m\right)\cr
	&-\Gamu mba\left(\partial_m Z_c - \Gamu ncm Z_n\right)\cr
	&-\Gamu mca\left(\partial_b Z_m - \Gamu nmb Z_n\right)\cr}$$
yielding
$$\eqalign{(\del a \del b - \del b \del a)Z_c
	&= (\Gamu mab - \Gamu mba)\del m Z_c
		+ (\partial_b\Gamu mca - \partial_a\Gamu mcb )Z_m \cr
	&~~~~~+ (\Gamu mca \Gamu nmb - \Gamu mcb \Gamu nma)Z_n \cr
	&= \tor mab \del m Z_c + \riemu ncba Z_n.\cr}\Eqno$$\label\RCompI
Rewriting equation \RCompI\ yields the correct abstract index expression for
the curvature tensor in the presence of torsion:
$$\riemu ncba Z_n = (\del a\del b - \del b\del a - \tor mab \del m) Z_c .
  \Eqno$$\label\RCompII

Thus, there is quite a difference between the treatment of torsion in the two
notational schemes.  While the first definition of curvature (equation \DefR )
proved to be valid with or without torsion, if one adopts the abstract index
notation to describe a theory involving torsion, one must also re-define the
curvature tensor to take this into account.
While the abstract index notation is usually used to describe torsion-free
theories ({\it e.g.,} general relativity), we will see that the presence of
``deficiency'' in a parametric theory of spacetime has analogous consequences.

We conclude this discussion of torsion by stating the symmetries of the
curvature tensor when torsion is present ({\it cf.}\
\Ref{{M. Spivak}, {\bf A Comprehensive Introduction to Differential Geometry,
vol.\ II}, {Publish or Perish, Inc., 1979}}\label\SPIVAK
).

\Theorem{$R$ and $T$ have the following symmetries:
\item{{\it i}.} $T(X,Y) = - T(Y,X)$
\item{{\it ii.}} $R(X,Y)Z = - R(Y,X)Z$
\item{{\it iii.}} $\metric{R(X,Y)Z}W = \metric{R(X,Y)W}Z$ if $\nabla$ is
compatible with $<\ ,\ >$.
\item{{\it iv.}} (the first Bianchi identity)
$$\eqalign{R(X,Y)Z &+ R(Y,Z)X + R(Z,X)Y\cr
	&= \del X T(Y,Z) + \del Y T(Z,X) + \del Z T(X,Y)\cr
	&+ T\left(X,\bracket YZ \right) + T\left(Y,\bracket ZX\right) +
T\left(Z,\bracket XY \right)\cr}\Eqno$$\label\RCyclic}
{Symmetries {\it i} and {\it ii} are immediate.  To show {\it iv} just
write out the cyclic sum, use the definition of $T$, and keep in mind the
Jacobi identity for bracket.  Explicitly we have,
$$\eqalign{R(X,Y)Z +& R(Y,Z)X + R(Z,X)Y \cr
	&= \del X(\del YZ - \del ZY) + \del Y (\del ZX - \del XZ) \cr
	&+ \del
Z(\del XY - \del YX)-\del{\bracket XY}Z - \del{\bracket YZ}X\cr
	& - \del{\bracket XZ}Y\cr
	&= \del X\left(T(Y,Z) + \bracket YZ\right) + \del Y\left(T(Z,X) + \bracket
XZ\right)  \cr
&+ \del Z\left(T(X,Y)+ \bracket XY\right)-\del{\bracket XY}Z - \del{\bracket
YZ}X \cr
	&- \del{\bracket XZ}Y\cr
&= \del XT(Y,Z) + \del Y T(Z,X) + \del Z T(X,Y)\cr
&+T(X,\bracket YZ) + T(Y,\bracket ZX) + T(Z,\bracket XY)\cr
&+\bracket X{\bracket YZ} + \bracket Y{\bracket ZX} + \bracket
Z{\bracket XY}\cr}$$
where the last three terms add to zero.
To prove {\it iii} we need to assume that $\nabla$ is compatible with the
metric $\langle\ ,\ \rangle$, thus writing
$$\eqalignno{\metric {\del X \del YZ}W &= X\metric{\del YZ}W - \metric{\del
YZ}{\del XW}\cr
&= X\metric{\del YZ}W - Y\metric Z{\del XW} + \metric Z{\del Y\del
XW}\cr
\noalign{\hbox{and}}
\metric{\del {\bracket XY}Z}W &= \bracket XY \metric ZW - \metric
Z{\del{\bracket XY} W}\cr
\noalign{\hbox{we have}}
\metric{R(X,Y)Z}W &=\metric{\del Y \del X W}Z - \metric{\del X \del YW}Z +
\metric {\del{\bracket XY}W}Z\cr
	&+X\metric{\del Y Z}W - Y\metric Z{\del X W} - Y\metric{\del XZ}W\cr
	&+ X\metric Z{\del YW} - \bracket XY \metric ZW\cr
	&=-\metric {R(X,Y)W}Z + XY\metric ZW - X\metric Z{\del Y W}\cr
	&- Y\metric Z{\del XW} - YX\metric ZW + Y\metric Z{\del XW}\cr
	&+X\metric Z{\del YW} - \bracket XY \metric ZW\cr
	&= - \metric{R(X,Y)W}Z .\cr}$$}


\Section{The Standard Gauss-Codazzi Formalism}

The Gauss-Codazzi equations relate the geometry of a manifold (with metric) to
the geometry of an embedded submanifold.  Specifically, the higher-dimensional
manifold induces a metric on the embedded surface, and thus gives rise to a
unique derivative operator (on the surface) and finally a curvature tensor.
The Gauss-Codazzi equations relate these induced quantities to the
higher-dimensional quantities.

Let $\Sigma$ be a (nondegenerate) hypersurface in $\M$, {\it i.e.} an embedded
submanifold of codimension 1 such that the metric $k$ induced on $\Sigma$ by
the metric $g$ on $\M$ is nondegenerate.  The induced metric is of course the
pullback of $g$ along the embedding, but it can also be expressed as a
projection operator as follows.

Let $n$ be the unit normal vector to $\Sigma$.
\Footnote{In the Lorentzian case, where ($\M$,$g$) is a {\it spacetime},
$\Sigma$ is typically a spacelike hypersurface, {\it i.e.}\ a Riemannian
manifold in its own right.  In this case, it is customary to choose $n$ to be
the {\it future-pointing} timelike unit vector field orthogonal to $\Sigma$.}
Then the induced metric is given by
$$k = g \pm n^\flat \otimes n^\flat$$
where $n^\flat$ is the 1-form dual (with respect to $g$) to $n$ and where the
sign depends on whether $n$ is timelike ($+$) or spacelike ($-$).

For any point $p\in \Sigma$, the tangent space $T_p\M$ may be written as a
direct sum
$$\eqalign{T_p\M &= T_p\Sigma \oplus (T_p\Sigma)^\perp\cr
	&=:\left(T_p\M\right)^\perp \oplus \left(T_p\M\right)^\top\cr}$$
where $\tpsigmaperp$ is the orthogonal complement of $T_p\Sigma$ in $T_p\M$
(with respect to the spacetime metric $g$).  For any $v\in T_p\M$, let $v^\top$
and $v^\perp$ be the obvious projections so that
	$$v = v^\perp + v^\top $$
where we have used $\perp$ to denote the projection to the {\it tangent} space
of $\Sigma$ (to agree with the notation of the next section).

Given vector fields $X$ and $Y$ on $\Sigma$, one may define a
connection on $\Sigma$ by
	$$D_XY = (\nabla_XY)^\bot .\Eqno$$\label\InducedDeriv
Equation \InducedDeriv\ not only defines an affine connection on $\Sigma$,
but, as is shown {\it e.g.}\ in \DOCARMO , $D$ is the unique Levi-Civita
connection associated with the induced metric $k$.
\Footnote{It is shown below that $D$ is torsion-free; metric compatibility
follows as a special case of Proposition 10 in the next section.}
One may define the curvature of $D$ in the usual manner:
$$\rthree(X,Y)Z = \d X \d Y Z - \d Y \d X Z - \d {[X,Y]} Z
\Eqno$$\label\RiemD
where $X,Y,Z$ are tangent to $\Sigma$.  Since $\Sigma$ is a
hypersurface, $\bracket XY$ denotes a vector field tangent to $\Sigma$ and,
hence, $\d {\bracket XY} Z$ is well-defined.  Using $\langle\ ,\ \rangle$ to
denote the spacetime metric, one can show, {\it e.g.}\ \DOCARMO , that the
curvature $\rfour$ of $\M$ and the curvature $\rthree$ of the surface
$\Sigma$ are related by Gauss' equation
	$$\eqalign{\left< \rfour(X,Y)Z, W\right> &=\left<\rthree(X,Y)Z,
W\right>\cr
	&-\left<B(Y,W), B(X,Z)\right> + \left<B(X,W),
B(Y,Z)\right>\cr}\Eqno$$\label\GaussEq
where all the vectors $X,Y,Z,W$ are assumed to be tangent to $\Sigma$ and
$B(X,Y)$ is the tensor defined by
$$\eqalign{B(X,Y) &= \del X Y - \d X Y\cr
	&=\left(\del XY\right)^\top.\cr}$$
Notice that $B(X,Y)$ is orthogonal to $\Sigma$.

\Theorem{Taking $\nabla$, $D$, and $B$ as defined above, if $\nabla$ is
torsion-free then
\item{{\it i.}} $D$ is torsion-free and
\item{{\it ii.}} $B$ is symmetric.}
{$$\eqalign{\tord(X,Y) & := \d XY - \d YX - \bracket XY \cr
	&= \left(\del XY\right)^\perp - \left(\del YX \right)^\perp - \bracket
		XY \cr
	&= \left(T\left(X,Y\right)\right)^\perp\cr
	&= 0}$$
since $\bracket XY^\perp=\bracket XY$ by Frobenius' theorem.  The symmetry of
$B$ follows from the torsion-free properties of both connections.  We have
$$\eqalign{B(X,Y) - B(Y,X)
	&= \del XY - \del YX - \left( \d XY - \d YX \right)\cr
	&= \Big( T(X,Y) + \bracket XY \Big) -
		\Big( \tord(X,Y) + \bracket XY \Big)\cr
	&= 0.}$$}

$B$ is closely related to the extrinsic curvature $K$ of $\Sigma$, which is
defined by
$$K(X,Y) = \metric{-\del XY}n .$$
The relationship between $K$ and $B$ is given by
$$\eqalign{K(X,Y) &= \metric{-\del XY}n\cr
	&=\metric{-B(X,Y)-\d XY}n\cr
	&=\metric{-B(X,Y)}n - \metric{\d XY}n\cr
	&=\metric{-B(X,Y)}n}$$
so that the symmetry of $K$ follows directly from the symmetry of $B$ when
$\nabla$ is torsion-free.  $B$ can be thought of as measuring the difference
between the geometries of $\M$ and $\Sigma$.  In fact, $B$ is identically zero
if (and only if) every geodesic of $\Sigma$ is also a geodesic of $\M$.

It is worth mentioning that the tensor $B$ fails to be symmetric if $\nabla$
possesses torsion.  If we let $\tornabla$ and $\tord$ represent the torsion
tensors associated with the respective connections $\nabla$ and $D$, then the
above calculation shows that
$$B(X,Y) - B(Y,X) = \tornabla(X,Y) - \tord(X,Y).$$
Therefore, the failure of $B$ to be symmetric is to be expected in the most
general setting.


\Section{A Generalized Gauss-Codazzi Formalism}

The above formalism lends itself nicely to the {\it slicing} viewpoint, in
which a manifold is foliated with (usually spacelike) hypersurfaces.  Both the
slicing and Gauss-Codazzi formalisms focus on decomposing the geometry into a
piece tangent to $\Sigma$ and a piece orthogonal to $\Sigma$.  One can view
these decompositions as a place to begin an initial value formulation; the
Gauss-Codazzi relations impose certain constraints on the initial data.

One may instead consider the {\it threading} viewpoint
\Ref{{R. Jantzen and P. Carini}, {\it Understanding Spacetime Splittings
and Their Relationships}, in {\bf Classical Mechanics and Relativity:
Relationship and Consistency}, ed.\ by G. Ferrarese, Bibliopolis, Naples,
185--241, 1991}\label\JANTZEN
, which is dual to slicing in that the manifold is now (regularly) foliated
with a (non-null) vector field.
\Footnote{A more complete discussion of the relationship between slicing and
threading appears in
\Ref{Stuart Boersma and Tevian Dray,
{\it Slicing, Threading \& Parametric Manifolds},
Gen.\ Rel.\ Grav.\ (submitted).}\label\PaperIII
.}
If this vector field is hypersurface
orthogonal, then the orthogonal hypersurfaces can be used as in the slicing
scenario.  But what happens if the vector field is {\bf not} hypersurface
orthogonal?

Given a non-null vector field $A$ (not necessarily unit), at each point $p$ in
$\M$ one still has the decomposition
$$T_p\M = \tpmperp \oplus (T_p\M)^\top.$$
For $v\in T_p\M$, write
	$$v =  v^\perp + v^\top$$
with $v^\perp$ orthogonal to $A$ and $v^\top$ parallel to $A$.  As
before, the spacetime metric induces
a metric $h$ on $\tpmperp$ defined by
	$$h = g - \frac {A^{\flat}\otimes
A^{\flat}}{\left<A^\flat ,A^\flat\right>}\Eqno$$\label\InducedMetric
where $A^\flat$ is the 1-form which is dual (with respect to the metric $g$)
to the vector field $A$.

Let $\perpvfs\subset\vfs$ denote the set of all vector fields (everywhere)
orthogonal to $A$.  For $X,Y \in \perpvfs$, one may define
the operator
	$$\d X Y = (\del X Y)^\perp.$$
\Proposition {$D$ satisfies the properties of an affine connection.
Specifically:
	$$\eqalign{1.\ &\d{fX + gY}Z = f\d X Z + g\d Y Z\cr
	2.\ &\d X (Y + Z) = \d XY + \d X Z\cr
	3.\ &\d X (fY) = f\d X Y + X(f) Y\cr}$$
for all vector fields $X,Y,Z\in \perpvfs$.}
{This is just a consequence of the linearity of projections.  First,
$$\eqalign{\d{fX +gY}Z&=\left(\del{fX +gY}Z\right)^\perp\cr
	&=\left(f\del XZ + g\del YZ\right)^\perp\cr
	&=f\d XZ + g\d YZ.\cr}$$
Second,
$$\eqalign{\d X(Y+Z) &=\left(\del XY + \del XZ\right)^\perp\cr
	&=\d XY + \d XZ.\cr}$$
Finally,
$$\eqalign{\d X(fY) &= \left(\del X fY\right)^\perp\cr
	&= \left( f\del XY + X(f) Y\right)^\perp\cr
	&=f\d XY + X(f)Y.\cr}$$
Therefore, $D$ is an affine connection.}

In the case where $\tmperp$ corresponded to the tangent space of some
hypersurface, it was stated that $D$ was the Levi-Civita connection of the
surface (with respect to the induced metric).   Although (in the present
scenario) $D$ is not, in general, the Levi-Civita connection on any
submanifold, we may
still investigate the familiar properties associated with the Levi-Civita
connection.  Using $\llbrak\ ,\ \rrbrak$ to represent the metric $h$, we have

\Proposition{If $\nabla$ is compatible with $g$, then $D$ is compatible with
the metric $h$.  That is,
$$X \Big( \llbrak Y,Z \rrbrak \Big) =
	\llbrak \d X Y,Z\rrbrak + \llbrak Y, \d X Z \rrbrak $$
for $X,Y,Z\in\perpvfs$.}
{For $X,Y\in\perpvfs$, we have $\llbrak X,Y\rrbrak = \left< X,Y\right>$.  Since
$\d XY =
\del XY - (\del XY)^\top$ and $\left< (\del XY)^\top ,Z\right> =0$, we have
$\metric {\d XY}Z = \metric {\del XY}Z.$  The fact
that $D$ is compatible with $h$ is now a consequence of the fact the $\nabla$
is
compatible with $g$.}

In the last section we showed that $D$ being torsion-free was an immediate
consequence of $\nabla$ being torsion-free.  In the present situation,
progress is hindered by the fact that while $\d XY - \d YX$ represents a
vector field orthogonal to $A$, $\bracket XY$ may not.  In fact, $\bracket XY
\in\tmperp$ for all $X$ and $Y$ in $\tmperp$ if and only if $\tmperp$ is
surface-forming (Frobenius' Theorem).  Thus, it is quite fruitless to compare
$\d XY - \d YX$ with $\bracket XY$.  One may, however, decompose $[X,Y]$ as
	$$[X,Y] = [X,Y]^\top + [X,Y]^\perp.$$
We may now measure the fact that $\tmperp$ is not surface-forming by the
existence of $\bracket XY^\top$ and use $\bracket XY^\perp$
to measure the torsion of $D$.
\Definition{The (generalized) torsion, $\tordperp$, associated with the
connection
$D$ is defined by
$$\tordperp (X,Y) = \d XY - \d YX - \bracket XY^\perp.$$}

\Lemma{The generalized torsion is precisely the projection of the torsion
associated with $\nabla$.}
{We have,
$$\eqalign{\tordperp (X,Y) &= \d XY - \d YX - \bracket XY^\perp\cr
	&=\Big(\del XY - \del YX - \bracket XY\Big)^\perp\cr
	&=T(X,Y)^\perp.\cr}$$
}

\Theorem{If $\nabla$ is torsion-free, then $\tordperp(X,Y) \equiv 0$ for all
$X,Y\in\perpvfs$.}
{$$\eqalign{\tordperp(X,Y)
	&= \Big(T\left(X,Y\right)\Big)^\perp\cr
	&= 0.}$$}
\noindent Therefore $D$ still inherits its (generalized) torsion only from
$\nabla$.

We will show below that, in a coordinate basis, the connection symbols,
$\Gamperpu ijk$, associated with $D$ obey the symmetry $\Gamperpu ijk =
\Gamperpu ikj$ if and only if $D$ is torsion-free ($\tordperp=0$).  Thus, the
above definition of $\tordperp$ is quite reasonable.

\Definition{The deficiency, $\D$, of the connection $D$ is defined by
$$\D(X,Y) = \bracket XY^\top.$$}

\Theorem{The following statements are equivalent:
\item{{\it i.}} $\tmperp$ is surface-forming.
\item{{\it ii.}} The generalized torsion $\tordperp$ associated with $D$ is
the (standard) torsion $\tord$ as defined by \DefT .
\item{{\it iii.}} $\D(X,Y)\equiv 0$ for all $X,Y\in\perpvfs$.}
{This theorem is basically the vector field version of Frobenius' theorem
rewritten to emphasize the new definitions.  By definition, $\D(X,Y)\equiv 0$
if and only if $\bracket XY^\top \equiv 0$.  Thus $\D(X,Y)\equiv 0$ if and
only if $\bracket XY\in\perpvfs$, yielding {\it iii} $\Leftrightarrow$ {\it i}
via Frobenius' theorem.  To show {\it iii} $\Rightarrow$ {\it ii}, we
again have $\bracket XY^\top \equiv 0$ so $\bracket XY^\perp \equiv \bracket
XY$, making the two notions of torsion coincide.  Since $\tordperp(X,Y) -
\tord (X,Y) = \bracket XY^\top$, we also easily have
{\it ii} $\Rightarrow$ {\it iii}.}

For $X,Y\in\perpvfs$, define as before
$$B(X,Y) = \del XY - \d XY .$$
$B(X,Y)$ is again a vector field orthogonal to the vector fields $X$ and $Y$
and is in fact tangent to $A$.  Even when $\nabla$ is
torsion-free, $B$ may still fail to be symmetric.
$$\eqalign{B(X,Y) - B(Y,X) &= \tornabla(X,Y) + \bracket XY
				- \tordperp(X,Y) - \bracket XY^\perp \cr
	&= \D(X,Y) + \tornabla(X,Y) - \tordperp(X,Y).\cr}\Eqno$$\label\GenB

\Theorem{If $\nabla$ is torsion-free, then $B(X,Y) = B(Y,X)$ if and only if
$\D(X,Y) = 0$.}
{$\tornabla(X,Y) = 0$ implies that $\tordperp(X,Y)=0$ and, hence, equation
\GenB\ reduces to
$$B(X,Y) - B(Y,X) = \D(X,Y).$$}

\noindent We have that the deficiency of the connection $D$ measures the
failure of $\tmperp$ to be surface-forming and, equivalently, the failure of
the extrinsic curvature $B$ to be symmetric in a torsion-free setting.

Being an affine connection, $D$ must have an associated ``curvature'' tensor.
However, the existence of the $[X,Y]^\top$ component prevents one from
proceeding as before --- ``$D_{[X,Y]}$'' doesn't make sense!  It appears as if
this problem may be overcome simply by using the quantity $[X,Y]^\perp$ to
represent the commutator of two vector fields orthogonal to the original
vector field $A$.  Armed with such a notion of ``bracket'', the next step
would be to define a curvature operator.
\Definition{Define the operator $S$ by
$$S(X,Y)Z = \d X \d Y Z - \d Y \d X Z - \d{[X,Y]^\perp} Z.\Eqno$$\label\SDef
}
Unfortunately, such a definition immediately leads to problems.

\Proposition{$S(X,Y)Z$ is not function linear (unless $\D=0$).  That is
$$S(X,fY)(gZ) \neq fg\, S(X,Y)Z.$$}
{$$\eqalign {S(X,fY)(gZ) &= \left( \d X (f\d Y) - f \d Y \d X -
\d{f[X,Y]^\perp} - \d{X(f)Y}\right) (gZ)\cr
	&= fS(X,Y)(gZ) + \Bigl( X(f)\d Y - X(f) \d Y\Bigr) (gZ)\cr
	&= fS(X,Y)(gZ)\cr
	&= f\Bigl( \d X \left(Y(g)Z + g\d YZ\right) - \d Y \left( X(g) Z + g
\d XZ\right)\cr
	&-[X,Y]^\perp (g) Z - g \d{[X,Y]^\perp} Z\Bigr)\cr
	&=f\left( [X,Y](g) Z - [X,Y]^\perp (g) Z + g S(S,Y)Z \right)\cr
	&= fg S(X,Y)Z + \D(X,Y) (g) Z \cr}$$
}

Therefore, in order to define a function linear curvature operator (tensor!),
we must keep track of the $[X,Y]^\top$ component (we can not just project it
away and forget about it).  That is, the $\d{[X,Y]^\perp} Z$ term in equation
\SDef\ is not complete.  We do not want to project the vector field $[X,Y]$
too soon!  We will, therefore,  consider replacing the last term of \SDef\ by
the term $(\del{[X,Y]}Z)^\perp$.  This
term is equivalent to the $\d{[X,Y]}Z$ term in equation \RiemD .  However,
since $[X,Y]$ is not necessarily orthogonal to $A$ we can not write
$(\del{[X,Y]}Z)^\perp$ in terms of the connection $D$.

\Definition{The (generalized) curvature operator associated with $D$ is
defined by
$$\rperp(X,Y)Z = \d X \d Y Z - \d Y \d XZ - (\del{[X,Y]}Z)^\perp.$$
}

\Proposition{$\rperp$ is function linear.  That is, $\rperp$ is tensorial.}
{$$\eqalign{
	\rperp (X,fY)&(gZ) =\cr
	&  \left( \d X (f \d Y) - f\d Y \d X - (\del{f[X,Y]}
+ \del{X(f)} Y)^\perp\right) (gZ)\cr
	&= f\rperp (X,Y)(gZ) + \left(X(f)\d Y - (\del{X(f)Y})^\perp\right)
(gZ)\cr
	&= f\rperp(X,Y)(gZ) + \Bigl( X(f)\d Y - X(f) \d Y\Bigr) (gZ)\cr
	&= f \rperp(X,Y)(gZ)\cr
	&= f\Bigl(\d X \left(Y(g)Z + g \d YZ\right) - \d Y \left(X(g) Z + g\d
X Z\right)\cr
	&-\left([X,Y](g) Z + g \del{[X,Y]}Z\right)^\perp\Bigr)\cr
	&=f\left( g\rperp(X,Y)Z + [X,Y](g)Z - ([X,Y](g) Z)^\perp\right)\cr
	&= fg \rperp (X,Y)Z\cr}$$
where the linearity of the projection map was used throughout.}

\Theorem{If $\nabla$ is metric compatible, then $\rperp$ satisfies Gauss'
Equation.  That is,
$$\eqalign{
\left<\rperp(X,Y)Z, W\right> = & \left<\rfour(X,Y)Z, W\right>\cr
	&+ \left<B(Y,W),B(X,Z)\right> - \left<B(X,W), B(Y,Z)\right>\cr
}\Eqno$$\label\GaussEqTwo
where $X,Y,Z$ and $W$ are orthogonal to $A$.}
{First, a few computational observations.  Since $B(X,Y) = \del X Y -
\d X Y$ is orthogonal to $A$,
$$\left< \del X Y , Z\right> = \left<\d X Y, Z\right>\Eqno$$\label\comment
for vector fields $X,Y,Z$ orthogonal to $A$.  While we have shown that $D$ is
compatible with the metric $\llbrak\ ,\ \rrbrak$, it is also true that, since
the full metric $\langle\ ,\ \rangle$ agrees with the induced metric
$\llbrak\ , \ \rrbrak$ on $\perpvfs$, one may write
$$X\Big(\left<Y,Z\right>\Big) =
	\left< \d X Y ,Z\right> + \left< Y, \d X Z \right>.$$
$D$ is thus ``compatible'' with the metric $\left<\ ,\ \right>$ when
restricted to the subspace $\perpvfs$.
Using the definition of $\rfour$ and $B$, we expand the right hand side of
equation \GaussEqTwo
$$\eqalign{RHS&= \left< \del X \del Y Z - \del Y \del X Z - \del{[X,Y]} Z ,
W\right>\cr
	&- \left<\del Y W - \d Y W , \del X Z - \d X Z\right>\cr
	& + \left< \del XW - \d X W , \del Y Z - \d Y Z\right>\cr
	&= X\left< \del Y Z,W\right> - \left< \del Y Z ,\del X W\right> -
Y\left<\del X Z ,W\right> \cr
	&+ \left< \del X Z , \del Y W\right> -
\left<\del{[X,Y]}Z , W\right> -\left<\del Y W , \del X Z\right>\cr
	& + \left< \del Y W , \d X Z\right> +
\left< \d Y W , \del X Z\right> - \left<\d Y W, \d X Z\right>\cr
	&+ \left<\del X W , \del Y Z\right> - \left< \del X W ,\d Y Z \right>
- \left< \d X W, \del Y Z\right> \cr
	&+ \left< \d X W ,\d Y Z\right>\cr
	&=X\left<\del Y Z ,W\right> - Y\left< \del X Z ,W\right> - \left<
\del{[X,Y]} Z ,W\right>\cr
	&+\left<\d Y W, \d X Z\right> - \left< \d X W, \d Y Z\right>\cr
	&=X\left<\d Y Z, W\right> - \left<\d Y Z ,\d X W\right>\cr
	&- Y\left<\d X Z ,W\right> + \left< \d Y W, \d X Z\right> -
\left<\left(\del{[X,Y]}Z\right)^\perp , W\right>\cr
	&=\left<\d X \d Y Z - \d Y \d X Z - \left(\del{[X,Y]}Z\right)^\perp
,W\right>\cr
	&=\left<\rperp (X,Y)Z, W\right>\cr}$$
where the second step involved the symmetry of the metric as well as equation
\comment .}

The above derivation of Gauss' equation only used the properties of metric
compatibility (for both pairs of connections and metrics).  In particular, the
symmetry (torsion) of either connection was not a concern. Thus, we have
further shown that Gauss' equation is valid in the presence of torsion.

Given $\langle\ ,\ \rangle$, $\rfour$, and $B$, one may use Gauss' equation to
{\it define}
a curvature operator $\rthree$.  In this context, we may view $\rperp$ as the
unique curvature tensor associated with $D$ which satisfies Gauss' equation.

A word of caution is necessary at this point.  If torsion is present in either
(or both) of the connections, the tensor $B(X,Y)$ will no longer be symmetric.
This affects the
symmetries of the tensors $\rperp$ and $\rfour$.  In particular,
as we shall see, $\rperp$ may not enjoy the familiar cyclic symmetry
$$\rperp(X,Y)Z + \rperp(Y,Z)X + \rperp(Z,X)Y = 0$$
even if $\rfour$ does!  However, the other symmetries are immediate.  More
precisely,

\Theorem {Let $\nabla$ be a torsion-free Riemannian connection
associated with the metric $\langle\ ,\ \rangle$, with curvature tensor $R$.
Using $D,
\llbrak \ ,\ \rrbrak$, $B$, and $\D$ as defined above, if $\rhat$ is an induced
curvature operator associated with the connection $D$ and $\rhat$ and $R$
satisfy Gauss' equation, then $\rhat$ has the following symmetries:
\item{{\it i.}}$\left< \rhat (X,Y)Z,W\right> = - \left<\rhat (Y,X)Z,W\right>$
\item{{\it ii.}}$\left<\rhat (X,Y)Z,W\right> = - \left<\rhat (X,Y)W,Z\right>$
\item{{\it iii.}}(first Bianchi identity)$$\eqalign{
\langle\rhat(X,Y)Z +& \rhat(Y,Z)X + \rhat (Z,X)Y,W\rangle \cr
	&=\left<B(X,W), \D(Y,Z)\right> + \left<B(Y,W) , \D (Z,X)\right> \cr
	&+ \left<B(Z,W), \D(X,Y)\right>\cr
	&=-\metric{\del X{\D(Y,Z)}}W - \metric{\del Y\D(Z,X)}W \cr
	& - \metric{\del Z \D(X,Y)}W\cr}.\Eqno$$\label\RCyclicII
}
{The symmetries in {\it i} and {\it ii} can be read off directly from equation
\GaussEq,\   keeping in mind that $R$ satisfies all of the symmetries of the
usual Riemann curvature tensor (in the absence of torsion).  To prove (iii),
just cyclicly permute $X,Y,$ and $Z$ in the terms on the right hand side of
equation \GaussEq\ and add, obtaining
$$\eqalign{\langle\rhat(X,Y)Z + &\rhat(Y,Z)X + \rhat
(Z,X)Y,W\rangle\cr
	&=\, 0-\left<B(Y,W), B(X,Z)\right> - \left<B(Z,W),
B(Y,X)\right>\cr
	& -\left<B(X,W), B(Z,Y)\right> + \left<B(X,W),B(Y,Z)\right>\cr
	& +\left<B(Y,W), B(Z,X)\right>+ \left<B(Z,W), B(X,Y)\right>\cr
	&=\left<B(X,W), \D(Y,Z)\right> + \left<B(Y,W), \D(Z,X)\right>\cr
	& + \left<B(Z,W), \D(X,Y)\right>.\cr}$$
which is the first line in {\it iii}.  However, this cyclic sum involving
$B$ and $\D$ may be rewritten in terms of $\nabla$ and $\D$.  Thus
written, claim {\it iii} resembles the standard cyclic symmetry of $R$ (see
equation \RCyclic ).  Keep in mind, however, that neither $\nabla$ nor $D$
possess torsion, although deficiency is present.  We have
$$\eqalign{\metric{B(X,W)}{\D(Y,Z)}&=\metric{\del XW - \d XW}{\D(Y,Z)}\cr
	&=\metric{\del XW}{\D(Y,Z)}\cr
	&=X\metric W{\D(Y,Z)} - \metric W{\del X{\D(Y,Z)}}\cr
	&= -\metric{\del X \D(Y,Z)}W\cr}$$
since $\D(Y,Z)$ is orthogonal to $W$.  Thus the second equation in {\it iii}
is true.}
Note that $W$ is arbitrary in {\it i} and {\it iii}, so that these can be
rewritten in the same form as Theorem 7 (except of course for the intermediate
result in \RCyclicII).

One further comment on the similarities between equations \RCyclic\ and
\RCyclicII\ is worth making.  In equation \RCyclic\ there are three extra terms
of the form $T\left(X,\bracket YZ\right)$ (and cyclic permutations).  One
might expect analogous terms in equation \RCyclicII\ involving $\D(X,\bracket
YZ ^\perp)$ and cyclic permutations.  However, since $\D(X,Y)$ represents a
vector field orthogonal to $A$,
$$\metric{\D\left(X,\bracket YZ^\perp\right)}W =0.$$
We see that the new concept of deficiency does indeed appear in the first
Bianchi identity in just the way torsion would.

\Section{Coordinate Expressions}

Let us now work in a coordinate patch and investigate the components of the
above operators.  For simplicity, we will consider the coordinate system
inherited from a threading decomposition of spacetime.
\Footnote{A more complete discussion of threading and its relationship to
parametric manifolds appears in \PaperIII.}
Let $A$ be timelike (and $\Sigma$ spacelike) with norm $1/M$, {\it i.e.}\
$\left<A,A\right>=-1/M^2$.   We now introduce coordinates
$x^\alpha = (x^0, x^i) = (t, x^i)$ such that the given vector field $A$ can be
written $A^\alpha = \frac 1{M^2}\left(\dt\right)^\alpha$.  $M$ is the
{\it threading lapse}; note that $A^0 = \frac 1{M^2}$ and $A^i=0$.  The
coordinates $x^i$ are constant along integral curves of $\dt$ and can thus be
thought of as coordinates on the (local) surfaces $\{ t\equiv {\rm
constant}\}$.  We assume throughout that a (Lorentzian) metric $g$ is given,
that $\nabla$ is its associated Levi-Civita connection, and that all other
tensors are as defined in the previous section.

Letting $m$ be the metric dual of the unit vector $A/\langle A,A \rangle$,
the {\it threading shift 1-form} is given by
$$ M_i \, dx^i := dt + \frac 1M m $$
Thus,
$$A_0 = -1 \qquad{\rm and}\qquad A_i = M_i.$$

In these coordinates, the spacetime metric $g$ takes the form
$$(g_{\alpha\beta}) = \left(\matrix{-M^2&M^2M_j\cr
\noalign{\hbox{\strut}}
 M^2M_i &
h_{ij}-M^2M_iM_j\cr}\right)$$

The functions $\hij = \gij + M^2 M_i M_j$ correspond to the components of the
{\it threading metric}, the metric on $\perpvfs$ induced by $g$ (equation
\InducedMetric).  These functions can also be thought of as the nonzero
components of the tensor
$$h_{\alpha\beta} = g_{\alpha\beta} + M^2 A_\alpha A_\beta$$
which is associated with the projection operator
$$\proj\alpha\beta = h_\alpha^{\ \beta} =
	\delta_{\alpha}^{\ \beta} + M^2 A_\alpha A^\beta$$
where $\downkdelta\alpha\beta$ is the Kronecker delta symbol.
Being a projection operator
guarantees that $\proj\alpha\beta X^\alpha = X^\beta$ for $X\in\perpvfs$.
It is easy to show that a spacetime vector field $X=X^\alpha\dxalpha = X^0\dt +
X^i\dxi$ is orthogonal to $A$ if and only if $X^0 = M_iX^i$.

To simplify notation we will introduce a ``starry'' derivative notation in
all coordinate directions.  Define
$$\partial_{*\alpha} = \partialalpha + A_\alpha \partialt.$$
Notice that since $A_0 = -1$ and $A_i = M_i$, we have
$$\eqalignno{\partial_{*0} &=0\cr
\noalign{\hbox{and}}
	\spartiali &= \partiali + M_i \partialt .\cr}$$
and we will often write $\spartiali f$ as $f_{*i}$.

Let us work out the action of the connection $D$ in these coordinates.  Given
$X$ and $Y$ in $\perpvfs$, we defined
$$\eqalign{\d XY&= (\del XY)^\perp\cr
	&= \proj\gamma\alpha X^\beta \del\beta Y^\gamma \dxalpha\cr
	&=\proj\gamma\alpha\proj\beta\delta X^\beta\del\delta Y^\gamma
\dxalpha\cr
	&=X^\beta\proj\gamma\alpha\proj\beta\delta\left( Y^\gamma\comma\delta
+ \Gamu\gamma\mu\delta Y^\mu\right)\dxalpha\cr
	&=X^\beta\left(\proj\gamma\alpha\left(Y^\gamma\comma\beta + M^2
A_\beta A^\delta Y^\gamma\comma\delta\right) + \proj\gamma\alpha
\proj\beta\delta \Gamu\gamma\mu\delta Y^\mu\right)\dxalpha\cr
	&=X^\beta\left( \proj\gamma\alpha Y^\gamma\starry\beta +
\proj\gamma\alpha\proj\beta\delta\proj\nu\mu Y^\nu \Gamu
\gamma\mu\delta\right)\dxalpha\cr
	&= X^\beta\left(Y^\alpha\starry\beta + M^2 A_\gamma A^\alpha
Y^\gamma\starry\beta + \Gamperpu\alpha\nu\beta Y^\nu\right)\dxalpha\cr
	&=X^\beta\left(Y^\alpha\starry\beta + \left( \Gamperpu\alpha\nu\beta -
M^2 A^\alpha A_{\nu *\beta}\right) Y^\nu\right)\dxalpha\cr}$$
where we have defined the symbol $\Gamperpu\alpha\nu\beta$ by
$$\Gamperpu\alpha\nu\beta = \proj\gamma\alpha\proj\beta\delta\proj\nu\mu
\Gamu\gamma\mu\delta. $$

It can be show that the symbol $\Gamperpu\alpha\nu\beta$ behaves like a
projected tensor.  That is,
$$\Gamperpu 000 = \Gamperpu \alpha 00 = \Gamperpu \alpha 0\beta =\Gamperpu
\alpha \beta 0 = \Gamperpu 0\beta 0 =0$$
\noindent and
$$\Gamperpu 0\alpha\beta = A_i \Gamperpu i\alpha\beta.$$
Since $\spartiali = \partiali + M_i \dt$ is a basis for $\perpvfs$, we define
the components of projected tensors by evaluating them on this basis.  Using
Lemma 12, it immediately follows that the components of the torsion
$\tordperp$ of $D$ are
$$\eqalign{
  \tordperp{}^k{}_{ij} \spartialk &= \tordperp(\spartiali,\spartialj) \cr
	&= \left( \Gamperpu kij - \Gamperpu kji \right) \spartialk
  }$$
which also shows that a torsion-free projected connection is indeed symmetric
as claimed.

Since $\d XY$ is orthogonal to $A$, $\d XY$ is completely determined by its
components $(\d XY)^i$.  That is
$$\eqalign{
(\d XY)^\alpha \dxalpha &= M_i (\d XY)^i \dt + (\d XY)^i \dxi \cr
			&= (\d XY)^i \spartiali \cr
  }.$$
But we have shown above that
$$(\d XY)^i = X^j \left( Y^i\starry j + \Gamperpu ikj Y^k \right)$$
where we have used the facts that $Y^i\starry 0 \equiv 0$ for all $Y$ and $A^i
=0$.  The above formula for $(\d XY)^i$ corresponds exactly to the parametric
covariant derivative operator introduced by Perj\'es \PERJES.  After a long
but straightforward calculation, one may show that in the absence of torsion
the terms $\Gamperpu ijk$ may be written in a familiar form involving the
parametric derivative operator and the components of the induced metric
$\hij$, which again agrees with Perj\'es:
$$\Gamperpu ijk = \frac 12 h^{im}(h_{mj*k} + h_{mk*j} -
h_{jk*m}).\Eqno$$\label\GamPerp
This provides covariant confirmation that Perj\'es' parametric structure can
be induced by a projective geometry of spacetime.

Continuing our coordinate description, let us calculate the components of the
curvature tensor $\rperp$ defined earlier.  The components of $\rperp$ are
defined by
$$\eqalign{\rperp (\spartiali , \spartialj )\spartialk &= \rperpu lkij
\spartiall\cr
	&= \rperpu lkij (\partiall + M_l \dt).\cr}$$

Calculating the ``spatial'' components of $\d{\spartiali} \d{\spartialj}
\spartialk$, we find:
$$\eqalign{\left(\d\spartiali \d\spartialj \spartialk \right)^l &=
\left(\d\spartiali \left(\d\spartialj \spartialk\right)\right)^l\cr
	&=\spartiali\left(\d\spartialj \spartialk\right)^l + \Gamperpu lni
\left(\d\spartialj \spartialk\right)^n\cr
	&=\spartiali\left(\kdelta lk \starry j + \Gamperpu lmj \kdelta
mk\right)\cr
	&+\Gamperpu lni\left(\kdelta nk \starry j + \Gamperpu nmj\kdelta
mk\right)\cr
	&= \Gamperpu ljk\starry i + \Gamperpu lni \Gamperpu nkj
.\cr}\Eqno$$\label\CompDD
Also,
$$\eqalign{\left[ \spartiali , \spartialj\right] &= (M_{j*i} - M_{i*j}) \dt\cr
	&= \D_{ji} \dt\cr}$$
where we have introduced the notation $\D_{ji}=M_{j*i} - M_{i*j}$.  Therefore,
$$\eqalign{\del{[\spartiali , \spartialj ]} \spartialk &= \D_{ji} \left(
{\partial M_k\over\partial t} + \Gamu 000 M_k + \Gamu 0k0\right)\dt\cr
	&+ \D_{ji} \left( \Gamu l00 M_k + \Gamu
	lk0\right){\partiall}\cr}$$
thus yielding
$${\left(\del{[\spartiali , \spartialj ]} \spartialk\right)^\perp} =
\left(\left(M_{j*i} - M_{i*j}\right)\left(\Gamu lk0 + M_k \Gamu
l00\right)\right) \spartiall.$$
Writing everything out gives us
$$\eqalign{\rperpu lkij &= \Gamperpu lkj\starry i - \Gamperpu lki\starry j +
\Gamperpu lni
\Gamperpu nkj - \Gamperpu lnj \Gamperpu nki\cr
	&+ 2 \left(M_{j*i} - M_{i*j}\right)\left(\Gamu l00 M_k + \Gamu
lk0\right) \cr
	&= \Gamperpu lkj\starry i - \Gamperpu lki\starry j + \Gamperpu lni \Gamperpu
nkj -
\Gamperpu lnj \Gamperpu nki\cr
	&+\left(M_{j*i} - M_{i*j}\right) h^{lm}\left(M^2 M_{m*k} - M^2 M_{k*m}
+ \partialt\, h_{km}\right)\cr}\Eqno$$\label\RPerpComp
where the symbols $\Gamma$ were replaced by the equivalent expressions
involving the threading metric, lapse function, and shift 1-form.  As we see,
the components of $\rperp$ are not quite as nice as in the case where the
$\spartiali$ span a hypersurface.  The non-zero contribution of $[\spartiali ,
\spartialj ]$ continues to complicate matters.

\Section{Zel'manov curvature}

In his work on parametric manifolds, which provided much of the motivation for
this current work, Perj\'es \PERJES\ gives the following {\it definition} of
the {\it Zel'manov curvature}
\Footnote{A similar expression appears in \EIN.}
$$\left[\sdel k \sdel j - \sdel j \sdel k +\left(\omega_{j*k} -
\omega_{k*j}\right)\dt\right]X_i = \zelmanovu rijk X_r\Eqno$$\label\Zel
(compare \RCompII) where $\omega_i$ can be identified with the $M_i$ defined
in the previous section.  In components, this takes the form \PERJES
$$\zelmanovu lkij = \Gamperpu lkj\starry i - \Gamperpu lki\starry j +
\Gamperpu lni \Gamperpu njk - \Gamperpu lnj \Gamperpu nik .$$
The Zel'manov curvature thus does not contain the contribution from
$[\spartiali , \spartialj ]$.
If one wants to relate the parametric tensor $\zelmanovu ijkl$ to a spacetime
tensor, one must re-examine the story leading up to the definition of
$\rperp$.

It seemed most natural to define $\rperp$ with the $(\del{[X,Y]})^\perp$ term,
as this definition closely resembles the definition of the standard curvature
tensor.  However, consider the definition
$$\rbarperp(X,Y)Z = \d X \d Y Z - \d Y \d X Z -
(\l{[X,Y]}Z)^\perp\Eqno$$\label\Rbarperp
where {\it \$} denotes Lie differentiation.

The difference between the two curvature operators is
$$\rperp(X,Y)Z - \rbarperp(X,Y)Z = -(\del Z\bracket
XY)^\perp\Eqno$$\label\CurvDiff
(assuming $\nabla$ is torsion-free).
In light of our
earlier comments, we know that $\rbarperp$ does not satisfy Gauss' equation.
However, there are the following similarities between $\rperp$ and
$\rbarperp$.
$$\eqalignno{1.\ &\rperp(X,Y) f = \rbarperp(X,Y)f \qquad{\rm for\ } X,Y
\in\perpvfs ,\cr
\noalign{\hbox{and in the case where $\tmperp$ is
surface-forming one has that $[\spartiali , \spartialj]=0 $\ which implies}}
	2.\ & \rperp(\spartiali , \spartialj) = \rbarperp(\spartiali ,
\spartialj)\cr}$$
so the two tensors agree in this special case.

For the components of $\rbarperp$, we must calculate
$\left(\l{\bracket\spartiali\spartialj}\spartialk\right)^\perp$.  If $\nabla$
is torsion-free, the definition of {\it \$} yields
$$\eqalign{\left(\l{\bracket\spartiali\spartialj}\spartialk\right)^\perp&=\bracket{\bracket\spartiali\spartialj}\spartialk\cr
	&=0\cr}\Eqno$$\label\CompL
Thus, the nonzero term $[\spartiali ,\spartialj ]$ does not contribute to the
components of $\rbarperp$.

We now show that we have in fact defined the Zel'manov curvature.
\Theorem{$\rbarperp$ is the Zel'manov curvature.}
{Using equations \CompDD\ and \CompL , we have
$$\eqalign{\rbarperpu ijkl &= \Gamperpu ikl\starry i - \Gamperpu lki\starry j +
\Gamperpu
lni\Gamperpu nkj - \Gamperpu lnj \Gamperpu nki\cr
	&= \zelmanovu lijk.\cr}\Eqno$$\label\ZelComp
}

\Section{Discussion}

The generalized Gauss-Codazzi approach seems to have been successful in
defining a notion of projected connection $D$, with a corresponding notion of
torsion.  Moreover, $D$ was found to be torsion-free if $\nabla$ was
torsion-free.  Most importantly, the deficiency $\D$ was explicitly defined in
such a way as to make its relationship to the torsion tensor clear.  While
distinct from torsion, deficiency plays much the same role {\it e.g.}\ in the
first Bianchi identity.
\Footnote{While we have not yet checked explicitly, we expect
the same will be true for the Codazzi equation and the second Bianchi
identity.}

However, there is one very peculiar aspect of this formalism, namely the
absence of an embedded hypersurface $\Sigma$ to which to project!  We have
chosen to interpret the Gauss-Codazzi formalism as providing a pointwise
projection, so that one obtains tensor fields defined on all of $\M$, rather
than on a preferred hypersurface $\Sigma$.

If the foliations are suitably regular, the orthogonal hypersurfaces in the
standard setting will be diffeomorphic to each other.  Thus, all projected
tensors can be viewed as living on the same such hypersurface, but at
different ``times''.  This viewpoint lends itself well to initial value
problems.  But this viewpoint also carries over to our generalized framework,
the only change being that one must work on the {\it manifold of orbits} of
the foliation, which is diffeomorphic (but {\bf not} isometric) to any
hypersurface of constant ``time''.

A {\it parametric manifold} is, in this setting, the manifold of orbits, on
which there are 1-parameter families of projected tensor fields.  The
geometric nature of its construction, using orthogonal projection, ensures
that it is {\it reparameterization invariant}, {\it i.e.}\ that it is
invariant under the special coordinate transformations which relabel ``time''.
This notion of 1-parameter families of tensor fields together with a suitable
reparameterization invariance can be used to give an intrinsic description of
a parametric manifold, without using projections; this will be published
separately
\Ref{Stuart Boersma and Tevian Dray,
{\it Parametric Manifolds II: Intrinsic Approach},
J. Math.\ Phys.\ (submitted).}%
.{}

Finally, we note that we have two somewhat different candidates, $\rperp$ and
$Z$, for the curvature of our projected connection.  The difference between the
two involves the deficiency, and hence vanishes in the hypersurface-orthogonal
case.  More importantly, the difference also involves the lapse function $M$,
which relates our arbitrary parameter $t$ to arc length (proper ``time'')
along the given curves.  For a given problem, which of these notions of
curvature is ``correct'' may thus depend on whether a notion of distance
orthogonal to the manifold of orbits is appropriate.

This brings us to the use of the Gauss-Codazzi formalism in initial value
problems, especially in general relativity.  The initial value formulation of
Einstein's equations imposes (consequences of) the Gauss-Codazzi equations as
constraints (and then uses the Mainardi equations to determine the evolution).
We thus conjecture that our framework can be used to generalize the initial
value formulation for Einstein's equations to appropriate data on the manifold
of orbits.  We further conjecture that it is precisely the (generalized) Gauss
and Codazzi equations which will lead to the appropriate constraints.  This
would also provide some evidence in favor of $\rperp$, which does satisfy
Gauss' equation, rather than $Z$, which doesn't.  We are actively pursuing
these ideas.


\vfill\eject
\bigskip\leftline{\bf ACKNOWLEDGEMENTS}\nobreak

This work forms part of a dissertation submitted to Oregon State University
(by SB) in partial fulfillment of the requirements for the Ph.D.\ degree in
mathematics.  This work was partially funded by NSF grant PHY-9208494.

\References

\bye